\documentclass[english,12pt,a4paper,twoside]{article}
\usepackage{amsfonts,}
\usepackage{amsmath}
\usepackage{color}
\usepackage{stmaryrd}
\usepackage{setspace}
\usepackage{amssymb}
\usepackage{graphics}
\usepackage{babel}
\usepackage{enumerate}
\usepackage{euscript}
\usepackage{dsfont}
\usepackage{soul}
\usepackage{newlfont}
\usepackage{latexsym}
\usepackage{graphicx}
\usepackage{a4wide}
\usepackage{cite}
\usepackage{url}
\usepackage[babel=true]{csquotes}

\newtheorem{Ellsberg Urn}{Ellsberg Urn}

\begin{document}

\title{Dynamic consistency of expected utility under non-classical (quantum)
uncertainty}
\author{Danilov V.I.\thanks{%
Central Mathematic Economic Institute, Russian Academy of Sciences,
vdanilov43@mail.ru.}, Lambert-Mogiliansky A.\thanks{%
Paris Scool of Economics, alambert@pse.ens.fr}, and V. Vergopoulos\thanks{%
Universit\'{e} Paris 1 Panth\'{e}on-Sorbonne and Paris School of Economics,
vassili.vergopoulos@univ-paris1.fr}}
\date{\today }
\maketitle

\begin{abstract}
Quantum cognition in decision-making is a recent and rapidely growing field.
In this paper we develop an expected utility theory in a context of
non-classical (quantum) uncertainty. We replace the classical state space
with a Hilbert space which allows introducing the concept of quantum
lottery. Within that framework we formulate axioms on preferences over
quantum lotteries to establish a representation theorem. We show that
demanding the consistency of choice behavior conditional on new information
is equivalent to the von Neuman-L\"{u}ders postulate applied to beliefs. A
dynamically consistent quantum-like agent may violate dynamic recursive
consistency, however. {This feature suggests interesting applications in
behavioral economics as we illustrate in an example of persuasion.}
\end{abstract}

\section{Introduction}

Alternatives available in decision problems can often be analyzed in terms
of a variety of perspectives: a fur coat may be evaluated from an esthetical
point of view or from the point of view of animal suffering. A military
intervention in Syria can be evaluated from a geopolitical perspective or
from a humanitarian one. Another type of example relates to the consumption
of cigarettes: the immediate pleasure perspective contra the long term
health perspective. In order to assess an alternative we need to build a
representation of it, a "represented alternative" which is a mental construct%
\footnote{%
Kahneman and Tversky (2000) write \textquotedblleft the true objects of
evaluation are neither objects in the real world nor verbal descriptions of
those objects; they are mental representations\textquotedblright \ a
conception which they further write is entirely natural for cognitive
scientists (p. xiv).}. Standard decision theory postulates that we always
are able to combine any relevant perspectives into a synthetic and stable
representation of the alternatives. However, it is also a common place for
cognitive scientists\textbf{\ }that we face difficulties when building our
representation of a complex alternative. We consider the alternative from
different perspectives - one at a time. And most importantly we are not
always able to synthesize information from various perspectives into one
single coherent and stable representation of the alternative.\

In this paper we are interested in decision-making under uncertainty and we
want to capture the difficulties people show in combining all relevant
information by analogy with incompatible properties in Quantum Mechanics. To
many people it may appear unmotivated or artificial to turn to quantum
mechanics (QM) when investigating human behavioral phenomena. However, the
founders of QM, including Bohr and Heisenberg, were early to recognize an
essential similarity between the two fields:\footnote{%
In particular Bohr was influenced by the psychology and philosophy of
knowledge of Harald H\"{o}ffding (see Bohr 1971 and the Introduction in
Bitbol 2009 for an insightful discussion).} in both fields the object of
investigation cannot (always) be separated from the process of
investigation. QM and in particular its mathematical formalism was developed
to respond to a general epistemological challenge: how can one study an
object that is being modified by the measurement of its properties? This
provides legitimacy to the exploration of the value of the mathematical
formalism of QM in the study of human behavioral phenomena without reference
to Physics.\footnote{%
The human mind behaves in a wide array of weird manners. It is not the
weirdness of quantum mechanics that makes it an attractive toolbox, but the
fact that it is a most general paradigm for structural contextuality (i.e.,
non-separability between the object of and the operation of investigation).}
Of particular interest in our context is that this formalism allows
representing agents subject to the incapacity to simultaneously consider a
choice alternative from all relevant perspectives. For instance when
evaluating the "animal suffering" value of a fur coat, its esthetical
(subjective) value, that was well-determined in our decision-maker's mind
before considering animal suffering aspects, may become "blurred" i.e.,
uncertain.

The classical approach to decision-making under uncertainty e.g., in Savage
(1972) and Anscombe and Aumann (1963) builds on the notion of a state space $%
S$ of states of nature. Very roughly a representation of the world (a
belief) corresponds to a probability distribution on $S.$ And the changes in
beliefs following the acquisition of new information follow Bayes' rule
which can be given a behavioral foundation as shown in Ghirardato (2002) :
Bayesian updating secures dynamic consistency i.e., it secures that choices
based on updated preferences are consistent with ex-ante preferences defined
for the condition (event) that triggered updating. There exists however
massive evidence of violations of Bayes rule. One source of violations is
that measurements (in a broad sense) affect the object of measurement. Most
clearly this happens in quantum physics and it is the reason why some
properties may be incompatible. This is formally expressed in the
non-commutativity of measurement operations which induces a non-Bayesian
updating process. A related line of motivation appeals to the growing
interest for applications of elements of the mathematical formalism of
Quantum Mechanics to psychology, social sciences and in particular to
decision-making (see e.g., Brandenburger and La Mura (2015), Khrennikov
(2014) and Busemeyer and Bruza (2012) for an overview of the field). The
approach has shown successful in explaining a large variety of behavioral
anomalies in decision-making ranging from cognitive dissonance, preference
reversal, conjunction fallacy, disjunction effects to framing effects.

\textbf{In a recent book Akerlof and Schiller (2015) labelled a new term
"Phishing equilibrium" to express how markets systematically exploits the
manipulability of real consumers with far reaching implications for the
efficiency and welfare properties of free markets. In a last section we
suggest in a simple economic example that quantum indeterminacy of beliefs
implies a "manipulability" of economic agents much in line with Akerlof and
Schiller's empirical evidence as well as with their understanding of the
underlying psychological mechanism "Just change people's focus and one
changes the decisions they make" (p.173).}

In this article, we substitute the Boolean lattice of events with a more
general lattice of projectors in the Hilbert space as the suitable framework
for modelling decision-making. The notions are introduced progressively and
require no previous knowledge of Quantum Mechanics or Hilbert spaces. We
show that a natural definition of a quantum lottery allows for the
formulation of decision theoretical axioms similar to the classical ones
with one exception. We need axiom A0 that secures the stability of
preferences over lotteries defined over different perspectives (resolutions
of the state space). This axiom \textbf{(}that we labelled "no-framing"%
\textbf{)} is trivially satisfied in the classical world (all lotteries can
be expressed in a single finest partition(resolution) of the state space).
We next show that the von Neumann-L\"{u}ders projection postulate of Quantum
Mechanics used as an updating rule is both necessary and sufficient for
dynamic consistency of preferences. In our context the von Neumann-L\"{u}%
ders postulate arises from purely behavioral considerations that is from a
requirement of consistency applying to conditional (on new information)
preference relations. Interestingly, the specificity of non-classical
uncertainty (also referred to as "contextuality") is shown to imply a
failure of the so-called "recursive dynamic consistency" (a dynamic version
of the Savage's Sure Thing Principle). We use this result to show that
quantum indeterminacy of beliefs implies a significant "manipulability" of
consumers.

There exists a few earlier works addressing quantum probabilities in the
context of decision-making. These include Deutsch (1999), Pitowsky (2003),
Lehrer and Shmaya (2006), Danilov and Lambert-Mogiliansky (2010) and
Gyntelberg and Hansen (2012)). In particular Pitowsky writes about "betting
on quantum measurements" but he is not working with preference relations.
Interestingly, he formulates a rule saying that the probability for any
specific outcome is independent of the specific measurement that yields it
as one of its possible results. This rule is very much in line with our
axiom A0. Lehrer and Shmaya propose a subjective approach to quantum
probabilities but they do not work with quantum lotteries. Danilov and
Lambert-Mogiliansky develop an expected utility theory in a general
non-classical uncertainty context (ortho-modular lattices). A first
distinction with the present work is that instead of assuming the existence
of a certainty equivalent, we build on fundamentals which brings us closer
to the approach of von Neumann and Morgenstein and Aumann. We also adopt the
structure of the Hilbert space which allows addressing more general type of
lotteries.\footnote{\textbf{In Danilov and Lambert-Mogiliansky 2010, only
direct measurement (orthogonal resolution of the unit) were considered. In
the present work we also address "fuzzy" measurement by means of POVM
(positive operator valued measurements). In such a context an outcome is a
probability distribution over events. }} These steps are necessary to
develop the core contribution of the paper which is related to the dynamics
of beliefs and choices in response to new information in a non-classical
uncertainty environment. Gyltenberg and Hansen (2012) work with Hilbert
space to develop an expected utility theory with subjective events. Their
static setting shows similarities with ours. However their analysis appeals
to a large number of axioms - 12 where we have 5 - and most importantly they
do not address the issue of dynamic consistency.

The present work is a contribution to both decision theory and the
foundations of quantum cognition. We extend previous works in two
directions. First, we provide a complete characterization of expected
utility theory under non-classical (quantum) uncertainty: a concise
formulation of sufficient and necessary axioms in terms of preferences over
quantum lotteries. Most importantly, this construction allows for a
transparent characterization of dynamic consistency of choice behavior in
such an environment. Finally, we discuss the value of the approach for
economics and illustrate it with an example of "Phishing for Phools".

The paper proceeds as follows. First, we introduce the concept of quantum
lottery which gives the opportunity to define basic elements of the
mathematical formalism. In section 3 we provide a straightforward
construction and a complete characterization of preferences over quantum
lotteries satisfying some standard properties. We formulate the
corresponding axioms and derive our representation theorem.\textbf{\ }In
section 4 we address the issue of information updating and formulate our
central theorem of dynamic consistency. Thereafter we discuss the value of
our results in economics and end with some concluding remarks.

\section{Quantum lotteries}

We are interested in a decision-maker's preferences over what we call
quantum lotteries. In this section we define the notion of quantum or
Q-lottery. As for any lottery, the prize that the DM obtains depends on the
realization of some event which is the outcome of a measurement, it is an
uncertain payoff. And the lotteries described below (roulette, horse and
quantum lotteries) differ essentially in the type of measurement that is
being performed. Therefore we first need to clarify the meaning of
measurement and in particular of a quantum measurement. But we shall start
by reminding basic facts about roulette lotteries and so called `horse
lotteries'.\medskip

\emph{Roulette lotteries}

Hereafter we let $X$ denote a set of prizes. A \emph{roulette lottery} (with
prizes in $X$) is defined by a collection of prizes $x_{1},...,x_{r}$
together with the probabilities $p_{1},...,p_{r}$ ($p_{i}$ are non-negative
real numbers with $\sum_{i}p_{i}=1)$ for obtaining the corresponding prize.
Such a lottery can be written as the string $l=(x_{1},p_{1};...;x_{r},p_{r})$%
, but we prefer to write it as a formal sum $l=\sum_{i}x_{i}\otimes p_{i}$.
We could think of it in the following way: a measurement in the form of a
`roulette' is performed and gives an outcome in the set $\{1,...,r\}$. The
probability of outcome $i$ is $p_{i}$ and, depending on the outcome of this
`measurement', a prize $x_{i}$ is paid.

Such lotteries can be identified with (simple) probabilistic measures on the
set $X$. We denote by $\Delta (X)$ the set of such measures (or lotteries).
Under well-known conditions, von Neumann and Morgenstern obtained that the
utility of a lottery $l=\sum _i x_i \otimes p_i$ for the decision-maker DM
is given by a number $U(l)=\sum_{i}p_{i}u(x_{i})$. Here $u:X\rightarrow
\mathbb{R}$ is a `utility function' defined on the set $X$ of prizes.\medskip

\emph{Horse lotteries}

The next concept is that of a `horse lottery' (in the terminology of
Anscombe and Aumann) or `act' (in Savage's terminology). A \emph{horse
lottery} is a mapping $f:S\rightarrow X$ from the set $S$ of `states of
nature' to the set $X$ of prizes. A measurement is performed in the form of
a `horse race' and, depending on the result of this measurement, the
corresponding prize is paid.

Again under suitable conditions the utility of a horse lottery $f$ can be
written as $U(f)=\sum_{s}p_{s}u(f(s))$, where $u:X\rightarrow \mathbb{\
\mathbb{R}}$ is again a utility function, and $p$ is a (subjective)
probability measure on the set $S$. A considerable simplification of the
conditions was achieved by Anscombe and Aumann when taking roulette
lotteries as prizes. They define a horse lottery as a function $%
L:S\rightarrow \Delta (X)$. A measurement defines the state $s$ of nature,
after that a drawing of the lottery $L(s)$ performs which gives a resulting
prize.

In order to smoothly move over to quantum lotteries, it is convenient to
present horse lotteries slightly differently. We denote by $l(s,x)$ the
corresponding probabilities for realization of outcomes $x$ in the lottery $%
L(s)$. Now we can form the functions $L_{x}:S\rightarrow \mathbb{R}$ by the
rule $L_{x}(s)=l(s,x)$. And rewrite our horse lottery $L$ as $%
\sum_{x}x\otimes L_{x}$. Here the function $L_{x}$ can be understood as a
\emph{plausibility} (or as a \emph{potentiality}) of getting prize $x$.

Generally, a finite family $(L_{i},i\in I)$ of functions $L_{i}$ on a set $S$
is called a \emph{positive decomposition of unit} if all these functions $%
L_{i}$ are non-negative and their sum $\sum_{i}L_{i}$ is equal to the
function $1_{S}$ identically equal to 1. One can understand such a family as
a classical fuzzy measurement device with the set $I$ of outcomes; in a
state $s$ of nature this measurement gives the outcome $i$ with probability $%
L_{i}(s)$. If we associate a prize $x_{i}$ to outcome $i$, we obtain a horse
lottery $L=\sum_{i}x_{i}\otimes L_{i}$.\medskip

\emph{Quantum lotteries}

A quantum lottery is also a bet on the outcome of a measurement, but now a
quantum one. A measurement of some `observable' is performed, and, depending
on the result obtained, our DM receives some prize. To formalize the notion
of quantum measurement we have to modify the notion of a state space. The
set $S$ is replaced by some Hilbert space $H$. The notion of function on $S$
is replaced by the notion of Hermitian operator Below we give precise
definitions ({a reminder of elementary notions about Hilbert spaces is
provided in Appendix 1}). For now we only say that the main difference with
the classical state space model is that the Hilbert space model allows for
measurements that cannot be performed simultaneously i.e., they are
incompatible with each other. Therefore the performance of a measurement can
modify the state of the system. \medskip

\textit{Quantum measurement}

\emph{A}\  \emph{quantum measurement device} is modeled by a finite
collection $(P_{i},\ i\in I)$ of Hermitian operators such that

a) all $P_i$ are nonnegative, and

b) $\sum _i P_i=E$.

In Physics such a collection is called POVM (positive operator valued
measure); we prefer to speak about \emph{positive decomposition of unit}
(PDU). The elements of $I$ are the possible outcomes of the device; the
operators $P_{i}$ express the potentiality for realization of the outcome $i$
in a way similar to the functions $L_{x}$ for horse lotteries (see above).

We shall distinguish between two classes of measurements. The first one
consists of \emph{von Neumann measurements} (they are known also as direct
measurements, orthogonal measurements, first kind measurements, and
reproducible measurements). They are defined by the requirement that the $%
P_{i}$ are orthogonal each other, that is $P_{i}P_{j}=0$ for $i\neq j$. It
is easy to see that in this case all operators $P_{i}$ are projectors.
Conversely, it can be shown that if all $P_{i}$ are projectors, they are
orthogonal each other.

The second and broader class of measurements includes $(Q_{i},i\in I)$ such
that $Q_{i}$ commute with each other, that is $O_{i}Q_{j}=Q_{j}Q_{i}$ for
any $i,j\in I$. We call such measurements \emph{internally consistent.} It
is easy to see that orthogonal operators commute, so that von Neumann
measurements are internally consistent.

Examples of such measurements are `von Neumann measurements with noise'. We
make a von Neumann measurement $(P_{i},i\in I)$ (with orthogonal $P_{i}$)
and after having obtained an outcome $i$, we use a roulette lottery $l_{i}$
with values in a set $X$ to determine the final outcome $x$. Such a
measurement device is modeled by the collection $(Q_{x},\ x\in X)$, where $%
Q_{x}=\sum_{i}l_{i}(x)P_{i}$. Conversely, any internally consistent
measurement can be represented as von Neumann measurement with noise.

An example of more general measurements is provided by the notion of a\emph{%
\ compound measurement}. Suppose we have two von Neumann measurements
devices, $\mathcal{P}=(P_{i},\ i\in I$ and $\mathcal{Q}=(Q_{j},\ j\in J)$.
Then we can form the compound measurement $\mathcal{PQ}$ with the set of
outcomes $I\times J$: we perform first measurement $\mathcal{P}$, then
perform $\mathcal{Q}$ and write the obtained outcomes $(i,j)$. The
corresponding PDU is $(P_{i}Q_{j}P_{i},\ i\in I,\ j\in J)$. If $P_{i}$
commute with $Q_{j}$, $\mathcal{PQ}$ is von Neumann measurement as well.
However in the general case, when $\mathcal{P}$ and $\mathcal{Q}$ are
incompatible, the obtained measurement $\mathcal{PQ}$ is not von Neumann
measurement. This construction is one of the justifications for considering
non-orthogonal measurements. Another line of justification relates to the
possibility of defining mixtures and restrictions. \medskip

\textit{Quantum Lottery}

As we already wrote, a quantum lottery is a bet on the outcome of a quantum
measurement. More precisely, a Q-lottery is a pair made of a quantum
measurement device $\mathcal{P}=(P_{i},\ i\in I)$ (the \emph{base} of the
lottery) and the prizes associated with the corresponding outcomes $(x_{i},\
i\in I)$. We write such a Q-lottery as $\sum_{i}x_{i}\otimes P_{i}$.
Intuitively the measurement $\mathcal{P}$ is performed and depending on the
outcome $i$ that obtains, the agent receives prizes $x_{i}$. The set of
Q-lotteries is denoted as QL(H).

\emph{A} \emph{constant} Q-lottery is a lottery of the form $x\otimes E$; it
gives the prize $x$ with certainty.

Any Q-lottery $\sigma =\sum_{i}x_{i}\otimes P_{i}$ can be written in the
\emph{canonical form } $\sum_{x}x\otimes Q_{x}$, where $Q_{x}=%
\sum_{i,x_{i}=x}P_{i}$.\footnote{%
Although the set $X$ can be infinite only a finite number of $Q_{x}$ differ
from 0.} {Let $\mathbf{QL}_{c}(H,X)$ }be the set{\textbf{\ }of canonical
Q-lotteries (or simply $\mathbf{QL}_{c}(H)$ because the specification of the
set $X$ does not play an essential role).} Intuitively, the initial Q-lottery%
\textbf{\ }$\sigma $ and its canonical form only differ in the way we write
them and therefore can be considered as equivalent. Below we formulate this
equivalence as our `no-framing' axiom A0.

As it is the case for classical lotteries, we can define mixtures of
Q-lotteries{, but we restrict the mixture operation to canonical lotteries}.
Suppose that {we have two Q-lotteries} in the canonical form: $\sigma
=\sum_{x}x\otimes P_{x}$ and $\tau =\sum_{x}x\otimes Q_{x}$. Then we can
construct a new {canonical} Q-lottery as the \emph{mixture} of the two (with
weights $\alpha $ and $1-\alpha $) $\alpha \sigma +(1-\alpha )\tau
:=\sum_{x}x\otimes (\alpha P_{x}+(1-\alpha )Q_{x})$. The new lottery is
interpreted as follows. First you use a "roulette\textbf{\ }device" to
determine which one of $\sigma $ or $\tau $ will be played and thereafter
you play one of them.\footnote{%
In the quantum case we do not in general have the equivalence - as in the
classical case - with the alternative interpretation of the mixture: play
both lotteries and use the roulette device to select which outcome
determines the prize afterwards. This is because in the general case the two
{measurements implicit in the} lotteries need not be compatible.} {(For
instance, if $x^{\ast }\in X$ is an outcome never paid by $\tau $, then $%
Q_{x^{\ast }}=0$ and $\alpha \sigma +(1-\alpha )\tau $ pays $x^{\ast }$
under the event $\alpha P_{x^{\ast }}$.)} The set {$\mathbf{QL}_{c}(H,X)$ of
canonical Q-lotteries} is a convex space. We shall use this structure
intensively in what follows.

\section{Construction and characterization of `nice' preferences}

We are interested in preference relations over quantum lotteries that
satisfy some `natural' properties, familiar from von Neumann and
Morgenstern, Savage, Aumann and others. We call such preference relations
"nice". We start with a straightforward construction of nice preferences,
thereafter we formulate their properties (axioms), and finally we show that
these properties fully characterize nice preferences.

In order to construct preferences (in fact, to construct the utility of
Q-lotteries), we should specify two things. First, a \emph{utility function}
$u:X\rightarrow \mathbb{R}$. Second, a linear `\emph{belief' functional} $%
\beta : \mathbf{Herm} (H)\rightarrow \mathbb{R}$ which is

a) positive in the sense that $\beta \left( A\right) \geq 0$ for $A\geq 0,$

b) normalized in the sense that $\beta \left( E\right) =1.$

With these two ingredients, we can define the `$(u,\beta )$-utility' $%
U(\sigma )=U_{u,\beta }(\sigma )$ of any Q-lottery $\sigma
=\sum_{i}x_{i}\otimes P_{i}$ as
\begin{equation*}
U_{{}}(\sigma )=\sum_{i}u(x_{i})\beta (P_{i}),
\end{equation*}
For Q-lotteries $\sigma \ $and $\tau $ we set $\sigma \preceq \tau $ if $%
U(\sigma )\leq U(\tau )$. Below we list some `nice' properties A0-A4 that
this preference relation $\preceq $ on $\mathbf{QL}\left( H\right) $
possesses.\medskip

\emph{No framing}\medskip

\textbf{A0.} \emph{For any} {$\sigma, \tau \in \mathbf{QL}(H)$} \emph{\ with
respective canonical forms }$\sigma ^{\prime }$\emph{\ and }$\tau ^{\prime
},\  \sigma \preceq \tau $\emph{\ }$\Leftrightarrow \sigma ^{\prime }\preceq
\tau ^{\prime }.\ $\medskip

Axiom A0 follows from the fact that the utility of a Q-lottery $\sigma
=\sum_{i}x_{i}\otimes P_{i}$ is equal to the utility of its canonical form $%
\sum_{x}x\otimes (\sum_{i,x_{i}=x}P_{i})$. Indeed, the utility of the latter
is equal to $\sum_{x}u(x)\beta
(\sum_{i,x_{i}=x}P_{i})=\sum_{x}\sum_{i,x_{i}=x}u(x_{i})\beta
(P_{i})=\sum_{i}u(x_{i})\beta (P_{i})=U(\sigma )$.

This axiom is implicit in the Savage and Anscombe-Aumann frameworks.
However, in generalizations of these frameworks it must be imposed
explicitly see e.g., Cohen and Jaffray (1980). They formulate an axiom of
`non influence of formalization' very similar to our axiom A0. There are
also other works that reject that axiom in order to allow for framing
effects see Ahn and Ergin (2010). \medskip

\medskip

\emph{Weak order}\medskip

$\mathbf{A1.}$ \emph{The preference relation $\preceq$ is a weak order, that
is complete and transitive.}\medskip

This follows from its representation via the utility $U$. \medskip

The next following two properties assert that the preference relation is
consistent with a mixture structure on the set $\mathbf{QL}(H)$.\medskip

\emph{Independence}\medskip

$\mathbf{A2.}$ \emph{{Let $\sigma ,\tau ,\varphi \in \mathbf{QL}_c(H)$} be
Q-lotteries, and $\alpha \in \lbrack 0,1]$. If $\sigma \preceq \tau $ then $%
\alpha \sigma+(1-\alpha )\varphi \preceq \alpha \tau +(1-\alpha )\varphi $.}%
\medskip

{Property A2 follows from the linearity of $\beta $.}\medskip

\emph{Continuity}\medskip

$\mathbf{A3.}$ \emph{\ {Let $\sigma ,\tau ,\varphi \in \mathbf{QL}_c(H)$},
and $\sigma \prec \varphi \prec \tau $. Then there exists $\alpha$ and $%
\beta $ $(0<\alpha ,\beta<1)$ such that $\alpha \sigma +(1-\alpha )\tau
\prec \varphi$ and $\varphi \prec \beta \sigma +(1-\beta )\tau $.}\medskip

Indeed, since $U(\sigma )<U(\varphi )<U(\tau ),$ we have $U(\varphi )>\alpha
U(\sigma )+(1-\alpha )U(\tau )$ for some $\alpha \in (0,1)$ and $U(\varphi
)<\beta U(\sigma )+(1-\beta )U(\tau )$ for some $\beta \in (0,1)$. So $%
U(\varphi )>U(\alpha \sigma +(1-\alpha )\tau )$ and $U(\varphi )<U(\beta
\sigma +(1-\beta )\tau )$.\medskip

\emph{Monotonicity}\medskip

This property is the most subtle. It asserts, roughly speaking, that if we
in a lottery $\sigma =\sum_{i}x_{i}\otimes P_{i}$ replace the prizes $x_{i}$
with better ones then new lottery will be preferred to the initial one.
However, this formulation is too weak for our aims, and we formulate it in
stronger form. To do that we first note that we can define the canonical
form not only for `prize valued Q-lotteries' but also for `roulette valued
Q-lotteries', that is for expression of the form $\sum_{i}l_{i}\otimes P_{i}$%
, where $l_{i}=\sum_{x}x\otimes l_{i}(x)$\ are roulette lotteries. The
canonical form of this lottery is $\sum_{x}x\otimes Q_{x}$, where $%
Q_{x}=\sum_{i}l_{i}(x)P_{i}$.\ {Let $\mathbf{QL}(H,\Delta(X))$ denote the
set of all roulette-valued Q-lotteries. We use this to extend the preference
relation to $\mathbf{QL}(H,\Delta(X))$ as follows: for any $\sigma, \tau \in%
\mathbf{QL}(H,\Delta(X))$} with respective canonical form $\sigma ^{\prime }$
and $\tau ^{\prime },$ $\sigma \preceq \tau $ if and only if $\sigma
^{\prime }\preceq \tau ^{\prime }.$\ We next have to consider the derived
preference relation $\preceq _{\Delta }$ on the set of ordinary (roulette)
lotteries $\Delta (X)$. For roulette lotteries $l$ and $m$ we set $l\preceq
_{\Delta }m$ if $l\otimes E\preceq m\otimes E$. Here $l\otimes E$ denotes a
constant Q-lottery, getting with certainty the prize $l$, and similarly for $%
m\otimes E$.

Note that the preference $\preceq _{\Delta }$ on $\Delta (X)$ is represented
by the {affine extension of $u$ from $X$ to $\Delta(X)$, that we still
denote by $u$}. Indeed, we have {$U(l\otimes E)=U(\sum_{x}x\otimes
l(x)E)=\sum_{x}u(x)\beta (l(x)E)=\sum_{x}u(x)l(x)=u(l)$}, because $\beta
(E)=1$. \medskip

$\mathbf{A4.}$ \emph{Let $\sigma =\sum_{i}l_{i}\otimes P_{i}$ and $\tau
=\sum_{i}m_{i}\otimes P_{i}$ {be elements of $\mathbf{QL}(H,\Delta(X))$ with
the same base}. If $l_{i}\preceq _{\Delta }m_{i}$ for any $i\in I$ then $%
\sigma \preceq \tau $.}\medskip

This property can be considered as a strong version of the sure-thing
principle of Savage. It is a simple consequence of the formula {$U(\sigma
)=\sum_{i}u(l_{i})\beta (P_{i})$} and of the non-negativity of $\beta
(P_{i}) $.\medskip

\textbf{Definition 3.} A preference relation $\preceq $ on the set $\mathbf{%
QL}(H)$ of Q-lotteries is \emph{nice} if it has the properties $\mathbf{A0-A4%
}$ (or satisfies the axioms $\mathbf{A0-A4}$). \medskip

The discussion above shows that a preference relation $\preceq _{u,\beta }$
built with the help of a utility function $u$ on $X$ and a linear functional
$\beta $ on $\mathbf{Herm}(H)$ is nice. Our first result asserts that the
reverse is also true.\medskip

\textbf{Theorem 1.} {\emph{A preference relation $\preceq$ on $\mathbf{QL}%
(H) $ is nice if and only if there exist a function $u$ on } $X$ \emph{and a
positive and normalized linear functional} $\beta$ \emph{on $\mathbb{P}(H)$
such that $\preceq =\preceq _{u,\mu}$.}}

\emph{Moreover, if the preference $\preceq $ is not trivial (that is there
exist Q-lotteries $\sigma $ and $\tau $ such that $\sigma \prec \tau $) then
$\beta$ is unique and $u$ is unique up to a positive affine transformation.}%
\medskip

The proof is in {Appendix 2}. {We now briefly sketch this proof. First, fix
a measurement device $\mathcal{P}=(P_i$, $i\in I)$. In the first stage of
the proof, we use the primitive preference $\preceq$ on $\mathbf{QL}(H)$ to
construct another preference $\preceq_\mathcal{P}$ on the set $\mathbf{QL}_%
\mathcal{P}(H,\Delta(X))$ of roulette-valued Q-lotteries with base $\mathcal{%
P}$. Viewing each such Q-lottery in $\mathbf{QL}_\mathcal{P}(H,\Delta(X))$
as an Anscombe-Aumann act from $I$ to $\Delta(X)$, we invoke the
Anscombe-Aumann (1963) theorem, and obtain a utility function $u_\mathcal{P}$
and a probability vector $\beta_\mathcal{P}$ on $I$ providing together a
Subjective Expected Utility (SEU) representation of $\preceq_\mathcal{P}$.
In the second stage of the proof, we show that these various SEU
represetations are consistent with each other; that is, that the functions $%
u_\mathcal{P}$ are essentially independent of $\mathcal{P}$ and that the
probability vectors $\beta_\mathcal{P}$ arise from a single belief
functional $\beta$.}

{In a classical Anscombe-Aumann setup, it is straightforward to obtain the
consistency of the various SEU representations accross different partitions
of the state space because acts are directly defined as functions from the
state space to the the roulette space. But suppose now that an act is rather
a structure $(E_1,l_1;\ldots;E_n,l_n)$ where $(E_1,\ldots, E_n)$ is a
partition of the state space and $(l_1,\ldots, l_n)$ is a corresponding
collection of lotteries on $X$. Such an act induces a function $\sum_{i}%
\mathds{1}_{E_i}l_i$. Under an additional axiom requiring two acts inducing
the same function to be always indifferent, again we can easily obtain the
consistency of the various SEU representations accross the various
partitions of the state space. Comming back at our nonclassical setup, it
takes a very similar axiom, namely A0, to obtain the consistency of the SEU
representations $(u_\mathcal{P},\beta_\mathcal{P})$,accross the various PDU.
And from there the full representation stated in Theorem 1 easily follows.}
We next investigate some consequences of our result and some
reformulations.\medskip \medskip

\textit{The shadow operator of a Q-lottery}

Let $\preceq $ be a nice preference relation on $Q$-lotteries. We fix some {%
function $u$ from $X$ to $\mathds{R}$}. We call the \textit{shadow operator}
of a lottery $\sigma =\sum_{i}x_{i}\otimes P_{i}$ the Hermitian operator $%
Sh(\sigma )=Sh^{u}(\sigma )$ defined as follows
\begin{equation*}
Sh(\sigma )=\sum_{i}u(x_{i})P_{i}.
\end{equation*}%
We note that $\beta (Sh(\sigma ))=U(\sigma )$, and consequently the utility
of Q-lottery $\sigma $ depends only on its shadow operator. The notion of
shadow operator is the equivalent of the notion of utility profile in the
classical framework. And in the same way the expected utility of an act only
depends on its utility profile. From here we could completely forget about
Q-lotteries and discuss the utility of Hermitian operators expressed by the
functional $\beta $.\medskip

\emph{Trace and belief operator}

There exists a remarkably useful way of representing the belief functional $%
\beta $ by means of a (Hermitian) operator of belief. For that we shall make
extensive use of the concept of the trace of an operator (and precisely here
the finite dimensionality of $H$ becomes important). {A reminder of the
definition and properties of the trace is provided in Appendix 1}. {Of
particular value in our context are two properties: commutativity } {$%
\mathbf{Tr}(AB)=\mathbf{Tr}(BA)$} {\ and the fact that the trace of any
Hermitian operator is a real number. }\medskip

\textbf{Definition 4.} We call a \emph{belief operator} (or a \emph{%
cognitive state}) any nonnegative Hermitian operator with the trace equal to
1.\medskip

Given a belief operator $B$, we can define the functional $\beta $ on $%
\mathbf{Herm}\left( H\right) $, setting $\beta \left( A\right) =\mathbf{Tr}%
\left( AB\right) $ for any $A\in \mathbf{Herm}\left( H\right) $.\medskip

\textbf{Lemma 1.} 1) \emph{The functional $\beta $ takes real values};

2) \emph{$\beta \left( A\right) \geq 0$ for $A\geq 0;$}

3) \emph{$\beta \left( E\right) =1$.}\medskip

\textit{Proof}.\textbf{\ }See Lemma in Appendix 1. $\square $\medskip

As a consequence, we obtain that the functional $\beta $ built on the belief
operator $B$ is a belief functional. Moreover any belief functional has such
a form (for a unique belief operator $B$). In fact the formula $(A,B)=
\mathbf{Tr}\left( AB\right) $ gives a scalar product (and thereby also the
structure of an Euclidean space) on the real vector space $\mathbf{Herm}%
\left( H\right) $. Relying on well-known description of linear functionals
on Euclidean space, we have proved the following \medskip

\textbf{Theorem $1^{\prime }$.} {\emph{A preference relation $\preceq$ on $%
\mathbf{QL}(H)$ is nice if and only if there exist a utility function $%
u:X\rightarrow \mathbb{R}$ and a belief operator $B$ such that $\preceq $ is
represented by the function $\sigma \mapsto \mathbf{Tr}\left( Sh^{u}\left(
\sigma \right) B\right).$}}

{\emph{Moreover, if the preference $\preceq $ is not trivial (that is there
exist Q-lotteries $\sigma $ and $\tau $ such that $\sigma \prec \tau $) then
$B$ is unique and $u$ is unique up to a positive affine transformation.}}%
\medskip

{\textbf{Remark}}. In quantum physics such belief operators are called
`states' or `density operators'. We shall refer to them as simply `beliefs'
or `cognitive state' because they allow constructing subjective
probabilities in a most suitable way. Indeed, let $B$ be a belief operator;
then, for any event $P$, $\beta _{B}(P)=\mathbf{Tr}(PB)$ is the subjective
probability for event $P$ when the cognitive state is $B$.\medskip

\textbf{Example 1.} Assume that belief operator $B$ is given as the
projector $P=P_{e}$ on a one-dimensional subspace $\mathbb{C} e\subset H,$
generated by a vector $e$ of length 1 (that is $(e,e) =1$; in Physics such
operators are called \textit{pure states}). In other words, $P( x) =( x,e) e$%
. It is easy to check that, for any Hermitian operator $A$ (viewed as a
Q-lottery), its utility $U( A)$ is equal to $\mathbf{Tr}(AP)=( Ae,e)$. In
other words, the quadratic form $(Ae,e)$ gives the utility of lottery $A$
when the beliefs are represented by a pure state $e$.

Such beliefs correspond to a maximally precise (subjective) representation
of the state of measured system.\footnote{%
In the quantum context, a (pure) maximal information information state is
not a complete information as in the classical context. In a maximal
information state any measurement that generates new information leads to
the loss of some other previously known information: the state changes.}%
\medskip

\textbf{Example 2.} Let us now consider the opposite case, when DM has
beliefs represented by the operator $B=E/\dim H$. Such beliefs can be
interpreted as `\emph{uniform uncertainty}'; roughly speaking, the DM
assigns equal probability to all states. The expected value of a lottery $A$
under such beliefs is $U(A)=\mathbf{Tr}(AE)/\dim (H)=\mathbf{Tr}(A)/\dim (H)$
which corresponds to the arithmetic average of the eigenvalues of operator $%
A $.

Note that whatever the beliefs $B$, the expected value of a lottery-operator
$A$ equals to some convex combination of its eigenvalues.

\section{Updating}

Suppose that beliefs of a DM are given by a belief operator $B$. That is,
our decision-maker believes that the state of the measured quantum system is
$B$. Alternatively, her cognitive state is such that she assigns
probabilities to events according to $B$. \textbf{This means the following. We consider here events as subspaces of $H$, or as projectors. If $P$ is a projector then the probability of $P$ (at the belief $B$) is equal to $\text{prob}(P)=\mathbf{Tr}(PB)$. Since $P$ is a projector, we can rewritten the latter as $\mathbf{Tr}(PBP)$. }  In the next section we return to the
interpretation of $B$. But for now suppose that she receives information of
relevance for the lotteries. This information may concern the prizes or be
relative to the system, the measurements of which determines the outcomes
and the prizes. In the following we restrict ourselves to the case when the
value of the prizes does not change (and is given by a fixed utility
function $u$) and all new information concerns the measured system. For
instance, the DM (or someone else) performs some intermediate\textbf{\ }%
measurement and learns as a result of the measurement that some event $P$
occurred. It is almost obvious that her beliefs and preferences on
Q-lotteries should change, the question we ask in this section is how should
her preferences on quantum lotteries change after receiving that
information?\medskip

In Quantum Mechanics, it is simply postulated that states change in
accordance with the von Neumann-L\"{u}ders postulate. More precisely, a
system that was in state $B$ transits to the state $B^{\prime }=PBP/\mathbf{%
Tr}(PBP)$ as a result of performing a measurement that yields the event $P$.
The operator $PBP$ is Hermitian and non-negative ($(PBPv,v)=(BPv,Pv)\geq 0$
by force of the nonnegativity of $B$). Thus, $B^{\prime }$ is indeed a
state. Here, we need to clarify why $\mathbf{Tr}(PBP)$ is different from
zero so we are allowed to divide by this number. As a matter of fact, we
understand $\mathbf{Tr}(PBP)$ as the probability to discover event $P$ in
cognitive state $B$. Thus, by analogy with standard Bayesian updating, the
von Neumann-L\"{u}ders postulate focuses on cases where the state assigns a
positive probability to event $P$. If the trace $\mathbf{Tr}(PBP)$ were
equal to $0$, that would mean that something happened which has zero
probability, i.e. an event that is considered impossible under belief $B$.
That is, the beliefs of our \textsc{DM} captured by the state $B$ are simply
incorrect and she has to update them in a more fundamental way.

We want to show that in quantum decision theory beliefs change in the same
way. Clearly, we have to make some assumptions. In order to determine which
assumptions we need, we return for a minute to the behavior of a classical
decision-maker. She has preference between functions (acts) defined on the
set $S$ of states of nature; suppose that she learns in addition that the
true state lies in some subset $T\subseteq S$. It is quite natural to assume
that her new preference depends only on values of these functions on the
subset $T$. That is, only on the restriction of the various functions to $T$%
.\medskip

\textit{I. Information as\textbf{\ }events }\medskip

We want to proceed with Q-lotteries in a way analogous to the classical
case. We shall assume that the new information comes from the performance of
a projective measurement and that the obtained result informs that event $P$
has occurred. Here $P$ is a projector on subspace $W$. We have to define
what we mean with the \textquotedblleft restriction of a
Q-lottery\textquotedblright \ on a subspace $W$.

As a subspace of the Hilbert space $H$, $W$ is also a Hilbert space. Given
an Hermitian operator $A$ on $H$, one can consider the operator $PA$ \emph{%
as an operator on } $W$ ($v\mapsto PA(v)$ for $v\in W$). To avoid a
confusion, we denote operator $PA$, conceived as an operator on $W$, by $A|W$
and call it the \emph{restriction} of $A$ on $W$. First we note that $A|W$
is an Hermitian operator (as an operator on $W$, not on $H$). Indeed, if $%
y,z\in W$ (and hence $Py=y,\ Pz=z$) then
\begin{equation*}
(PAy,z)=(Ay,Pz)=(Ay,z)=(y,Az)=(Py,Az)=(y,PAz).
\end{equation*}%
The same argument shows that $A|W$ is a nonnegative operator on $W$ provided
$A$ is nonnegative. Note, finally, that $P|W$ is the identity operator on $W$%
.

The above indicates how to define a restriction to subspace $W$ of any
Q-lottery. If $\sigma =\sum_{i}x_{i}\otimes P_{i}$ is a Q-lottery on $H$,
then $\sigma |W:=\sum_{i}x_{i}\otimes P_{i}|W$ is a Q-lottery on $W$. We
call $\sigma |W$ the \emph{restriction} of lottery $\sigma $ to $W$.
Clearly, $Sh(\sigma |W)=Sh(\sigma )|W$.\medskip

Let us next turn to the problem of updating nice preferences. Suppose that $%
\preceq $ is a nice preference relation on the set $\mathbf{QL}(H)$; due to
Theorem $1^{\prime }$ it is given by some belief operator $B$. Suppose now
that our DM receives information in the form of an event-subspace $W\subset
H $ (or of an event-projector $P$). The updated preference relation will be
denoted as $\preceq _{W}$. Following Ghirardato (2002), we formulate two
axioms connecting $\preceq _{W}$ and $\preceq $. The first one A5 is
'consequentialism': \medskip

\textbf{A5}\emph{\ If }$\sigma $\emph{\ and }$\tau $\emph{\ are Q-lotteries
on }$H$\emph{, and }$\sigma |W=\tau |W$\emph{, then }$\sigma $\emph{\ and }$%
\tau $\emph{\ are equivalent with respect to }$\preceq _{W}$\emph{.}\medskip

To formulate the second axiom we need one more notion. We say that a
projector $P$ is compatible with a Q-lottery $\sigma =\sum_{i}x_{i}\otimes
P_{i}$ if $P$ commutes with every $P_{i}$. Now we can formulate axiom
\textbf{A6} of compatible dynamic consistency:\medskip

\textbf{A6} \emph{Suppose that lotteries }$\sigma $\emph{\ and }$\tau $\emph{%
\ are compatible with }$P$\emph{, and }$\sigma |W^{\perp }=\tau |W^{\perp }$%
\emph{, then }$\sigma \preceq _{W}\tau $\emph{\ if and only if }$\sigma
\preceq \tau $\emph{.\medskip }

Extending {Ghirardato (2002)} to the quantum context, we state the
following\medskip

\textbf{Theorem 2.} \emph{Let }$\preceq $\emph{\ be a non-trivial nice
preference given by a belief operator }$B$\emph{, and }$Tr(PBP)>0$\emph{.
Then}

\emph{a) The preference relation }$\preceq _{W}$\emph{\ given by the belief
operator }$B^{up}=PBP/Tr(PBP)$\emph{\ satisfies axioms \textbf{A5} and
\textbf{A6}.}

\emph{b) Conversely, if a preference relation }$\preceq _{W}$\emph{\
satisfies Axioms \textbf{A1}, \textbf{A5} and \textbf{A6}, then it is nice
and it is given by the `updated' belief operator }$B^{up}=PBP/Tr(PBP)$\emph{.%
}\medskip

The proof of Theorem 2 is in {Appendix 3}. {It generalizes the classical
equivalence result between Dynamic Consistency and Consequentialism on the
one hand and Bayesian updating on the other hand to our nonclassical setup.
Its proof is similar to the classical one. Given two classical Q-lotteries $%
\sigma,\tau \in \mathbf{QL}(H)$, we construct two other Q-lotteries $%
ad_W(\sigma),ad_W(\tau) \in \mathbf{QL}_c(H)$ with the following properties:
}

{(1) $ad_{W}(\sigma )|W=\sigma |W$ and $ad_{W}(\tau)|W=\tau |W$,}

{(2) $ad_{W}(\sigma )|W^{\perp }=ad_{W}(\tau)|W^{\perp }$.}

{That is, $ad_W(\sigma)$ and $ad_W(\tau)$ agree on $W$ with $\sigma$ and $%
\tau$ respectively, while they agree with each other on $W^{\perp }$. Given
Axioms A5 and A6, these two properties imply the equivalence between $\sigma
\preceq_W\tau$ and $ad_{W}(\sigma)\preceq ad_{W}(\tau)$ for any Q-lotteries $%
\sigma$ and $\tau$. In a classical setup, this equivalence is essentially a
form of Savage's (1954) Sure Thing Principle. Actually, Savage postulates a
preference satisfying the Sure Thing Principle and defines conditional
preference through this equivalence. Moreover, we show that we can choose $%
ad_W(\sigma)$ and $ad_W(\tau)$ such that their shadow operators are
respectively given by $PSh(\sigma)P$ and $PSh(\tau)P$. In classical terms,
this means that the `utility profiles' induced by $ad_W(\sigma)$ and $%
ad_W(\tau)$ are equal to those of $\sigma$ and $\tau$ on event $W$ and equal
to $0$ otherwise. From there, it takes a little algebra and the uniqueness
part of Theorem 1' to conclude.} We next discuss some corollaries of this
theorem.\medskip

\textbf{Remark 1.} We above assumed that the probability of the event $P$ is
non-zero. In the opposite case the received information contradicts the
initial beliefs. Consider now the case when the probability of the event $P$
(equal, as we know, to $\mathbf{Tr}(PBP)$) is 1, that is our DM is sure that
event $P$ must occur. Receiving information that $P$ occurred she does not
learn anything and intuitively we expect her preferences to remain the same,
which indeed obtains: \medskip

\textbf{Proposition 1.} \emph{A distance between a prior $B$ and the
posterior $B^{up}$ is} $O((1-\mathbf{Tr}(PBP))^{1/4})$.\medskip

In particular, if $Tr(PBP)=1$ then $B^{up}=B$. We prove Proposition 1 in
Appendix 4.\medskip

\textbf{Remark 2.} Let us consider a situation when the initial (a priori)
beliefs are maximally precise, that is, they are given (as in Example 1) by
a one-dimensional projector or, to put it differently, $B$ is a pure state.
How does the DM update her beliefs as she receives new information? We
expect the updated state to be pure as well. This is true in the classical
context and this is true in the quantum case as well with a noticeable
distinction that the new pure state is generally \textit{not} the same as
the initial one.\medskip

\textbf{Proposition 2.} \emph{If a belief operator $B$ is a one-dimensional
projector then the updated operator $B^{up}$ is a one-dimensional projector
as well. }\medskip

\textit{Proof.} The rank of operator $PBP$ is less or equal to the rank of
operator $B$, that is $\leq 1$. $\Box $\medskip

The main difference with the classical situation is that the new state is
generally different from the initial one. Indeed, let our belief operator $B$
be the projection on the line ${\mathbb{C}}e$, where $e$ is a normalized
vector ($(e,e)=1$). Such a projector $B$ moves an arbitrary vector $x$ to
vector $(x,e)e$. Then the operator $PBP$ moves vector $x$ to $%
P((Px,e)e)=(Px,e)P(e)=(x,Pe)Pe$, that is the projection (up to the
multiplier $(Pe,Pe)$) on the line ${\mathbb{C}}Pe$. Thus the pure state $e$
is changed into another pure state $Pe/\sqrt{(Pe,Pe)}$. The reader can see
here the projection postulate at work. We would like to add that the
modification is smaller the closer the vector $e$ to the subspace $W$
associated with the projector $P$. That is the closer $\mathbf{Tr}%
(PBP)=(Pe,Pe)$ to 1.\medskip

\textbf{Remark 3.} Suppose that, after receiving information in the form of
an event-projector $P,$ the DM receives yet some new piece of information in
the form of an event-projector $Q$ (which does not contradict $P$). Then (up
to a factor) the state changes as $B\mapsto PBP\mapsto QPBPQ$. However, in
the case when events-projectors do not commute with each other, the updated
beliefs depend on the order in which the updating is realized. Manipulating
the order in which information is provided affects our DM preferences and
thus her choice behavior. As we shall see later, together with Remark 2,
Remark 3 paves the ground for interesting economic implications. \medskip

\textbf{Remark 4.} Above we assume that the information comes in the form of
an event, that is a subspace $W$ or a projector $P$. However we could
consider the more general case when the information has the form a
`fuzzy-projector', that is an Hermitian operator $P$ such that $0\leq P\leq
E $. In this case the updating of the belief take the form $B^{up}=\sqrt{P}B%
\sqrt{P}/\mathbf{Tr}(PB)$. Proposition 1 generalizes to this setting; see
Appendix 4. \medskip \medskip

\emph{II. Information as measurements}\medskip

In the remaining of the paper we use the `shadow operator' expression for
lotteries: $A=Sh^{u}( \sigma ) $. Due to Theorem $1^{\prime }$, the utility
of such a lottery under belief $B$ is equal to $U\left( A\right) = \mathbf{Tr%
}(AB)$.

Usually, new information arises as an outcome of some measurement. Suppose
we perform some von Neumann measurement, represented by an orthogonal
decomposition of unit\textbf{, }ODU $(P_{i},\ i\in I)$. If, as the result of
this measurement, we obtain outcome $i$, the belief-state $B$ changes (is
updated) into $B_{i}=P_{i}DP_{i}/\mathbf{Tr}(P_{i}B)$. We say that the state
of the system \textit{transits} into subspace $W=\text{Im}(P_{i})$. Note
that the number $p_{i}=\mathbf{Tr}(P_{i}B)$ is precisely the probability
(under the belief $B$) for the realization of outcome $i$ when performing
our measurement. {Excluding impossible results, we can assume that these
numbers are stricty positive.}

The utility of `lottery' $A$, after the DM received information about
realization of outcome $i$, is now equal to $U_{i}\left( A\right) =\mathbf{Tr%
}(AB_{i})$. It may clearly be either larger or smaller than the initial $%
U(A) $.

Further we discuss an interesting case when such an `informational'
measurement has been performed (using ODU $\mathcal{P}=(P_{i},\ i\in I)$),
but its outcome is not known to the DM, i.e., she only knows that the
measurement took place. This is a new type of information; in Physics one
speaks about `decoherence'. It is a fully non-classical phenomenon, because
in the classical situation such an `information' is useless. This is not the
case in a quantum context.

We earlier established that when learning about the occurrence of an outcome
$i,$ the utility of the lottery-operator $A$ is updated to $U_{i}(A)=\mathbf{%
Tr}(AB_{i})$ (where $B$ is the belief of our DM, and $B_{i}=P_{i}BP_{i}/%
\mathbf{Tr}(P_{i}B)$) that is equal to $\mathbf{Tr}(AP_{i}BP_{i})/p_{i}$.
Since the probability of outcome $i$ is $p_{i}$, then the expected utility
of our lottery $U^{\prime }(A)$ (when learning about the performance of
measurement $\mathcal{P}$) is
\begin{equation*}
U^{\prime }(A)=\sum_{i}p_{i}U_{i}(A)=\sum_{i}\mathbf{Tr}(AP_{i}BP_{i}).
\end{equation*}%
(Of course, this means also that the state of belief $B$ has changed into a
new state $B^{\prime }=\sum_{i}P_{i}BP_{i}$). And, as we shall see, the
connection between these two utilities, i.e. \textit{ex-ante} $U(A)$ and
\textit{ex-post} $U^{\prime }(A)$, is not straightforward in general.
Ex-post utility can be either larger or smaller than the initial (a priori)
utility as the following simple example shows.\medskip

\textbf{Example 3.} Let $H$ be a two dimensional Hilbert space with
orthonormal basis $(e_{1},e_{2}).$ Let $A$ be a projector on $e_{1}$, i.e.,
an operator of the form {\footnotesize $A=
\begin{pmatrix}
1 & 0 \\
0 & 0%
\end{pmatrix}%
.$} Consider $A$ as a lottery that gives a utility equal to 1\ in\ state $%
e_{1}$ and 0 otherwise. Consider another lottery-operator {\footnotesize $C=
\begin{pmatrix}
0 & 0 \\
0 & 1%
\end{pmatrix}%
$} that gives utility 1 in state $e_{2}$ and 0 otherwise. Assume now that
our DM's belief-state is given by $B=A$.\ Then clearly the expected utility
of lottery $A$ is equal to 1 and the expected utility of $C$ is equal to
zero. So our DM strictly prefers $A$ to $C$.

Assume now that we perform a measurement defined by the ODU $(P_{1},P_{2})$,
where {\footnotesize $P_{1}=
\begin{pmatrix}
1/2 & 1/2 \\
1/2 & 1/2%
\end{pmatrix}%
$} and {\footnotesize $P_{2}=
\begin{pmatrix}
1/2 & -1/2 \\
-1/2 & 1/2%
\end{pmatrix}%
.$} If the outcome of the measurement is 1, the updated belief-state is
given by operator $B_{1}=P_{1}BP_{1}/\mathbf{Tr}\left( P_{1}BP_{1}\right) ,$
which as can be seen easily is equal to $P_{1}.$ The expected utility in the
belief-state is $U_{1}\left( A\right) =\mathbf{Tr}(AB_{1})=\mathbf{Tr}%
(AP_{1})=1/2.$ And similarly if we obtain the complementary result $2,$\ the
belief-state is updated to $B_{2}=P_{2}BP_{2}/\mathbf{Tr}(P_{2}BP_{2})$ and
the corresponding expected utility is $U_{2}\left( A\right) =\mathbf{Tr}%
(AB_{2})=\mathbf{Tr}(AP_{2})=1/2$. So we see that for any one of the two
possible outcomes the value of the $A$ lottery goes from 1 to 1/2. With the
same reasoning we obtain that $U_{1}\left( B\right) =\mathbf{Tr}(CB_{1})=%
\mathbf{Tr}(CP_{1})=1/2=U_{1}\left( C\right) .$ So the two lotteries $A$ and
$C$ yield the same expected utility. This violates "recursive dynamic
consistency": lottery $A$ is \textit{ex-post} indifferent to $C$ whatever
the outcome of the measurement, yet \textit{ex-ante} it is strictly
preferred. Note that the intermediary measurement $\left( P_{1},P_{2}\right)
$ is incompatible with either $A$ or $C$ and $B$. \medskip

It may seem odd that our dynamic consistency axiom A6 allows for such
departures. But the appeal of recursive dynamic consistency is based upon
the implicit assumption that "the act the agent performs has no effect on
the resolution of uncertainty" (cf. Fishburn, 1970). However, the resolution
of uncertainty is - in our setting - affected by the act that is selected
and the measurements it entails (as well as by other measurements) performed
to acquire new information. Once this is taken into account, recursive
dynamic consistency loses much of its appeal. In the next section we discuss
an economic example of the phenomenon exhibited above and establish its
connection with the behavior documented in the "Economics of Manipulation
and Deception" (Akerlof and Schiller, 2015). \medskip

Example 3 allows illustrating `information as measurements' a feature that
lacks counter-part in the classical model. Imagine that we perform the
measurement described above but our DM is not informed of the result. She
only knows the measurement has been made. In the classical world such an
information does not affect the DM's belief or the expected value of the
lotteries. However in the quantum context the DM understands that for any of
the two outcome (1 or 2) the expected value of lottery $A$ has changed from
1 to 1/2. Therefore, independently of her (lack of) knowledge about the
outcome of the measurement, she will revise her belief-state only because
she knows that this specific measurement has been performed. The new
belief-state is $B^{\prime }=p_{1}B_{1}+p_{2}B_{2}=E/2$ which corresponds to
`uniform ignorance'. And in this belief-state the expected utility of
lottery $A$ is equal to $\mathbf{Tr}(AB^{\prime })=\mathbf{Tr}(AE)/2=1/2$. $%
\Box $\medskip

There are two interesting cases when decoherence (that is a measurement with
unknown outcome) does not change the utility of a lottery or the state of
belief. These are cases when the measurement $\mathcal{P}$ is compatible
either with the lottery $A$ or with the belief $B$, i.e. when all the
projectors $P_{i},\ i\in I$, commute either with operator $A$ or with
operator $B$.\medskip

\textbf{Proposition 3.} \emph{Assume that the intermediate measurement is
compatible with either operator A or B. Then $U(A)=U^{\prime }(A)$.}\medskip

\textit{Proof.} $U^{\prime }(A)=\sum_i \mathbf{Tr}(AP_iBP_i)=\sum_i \mathbf{%
Tr}(P_iAP_iB)$. Suppose that $P_iA=AP_i$. Then the second sum can be
rewritten as
\begin{equation*}
\sum \mathbf{Tr}(P_iP_iAB)=\sum \mathbf{Tr}(P_iAB)=\mathbf{Tr}((\sum P_i)AB)=%
\mathbf{\mathbf{Tr}}(AB)=U(A).
\end{equation*}
Suppose now that $P_iB=BP_i$. Then the first sum can be rewritten as
\begin{equation*}
\sum \mathbf{Tr}(AP_iP_iB)=\sum \mathbf{Tr}(AP_iB)=\mathbf{Tr}(A(\sum P_i)B)=%
\mathbf{Tr}(AB)=U(A). \  \  \blacksquare
\end{equation*}

\textbf{Remark 5.} Decoherence always changes state of belief toward a more
dispersed one. One can give to this statement an exact sense, using notions
from the book of Alberti and Uhlmann "Stochasticity and partial order",
1982. Here we would like to illustrate this by an example when initial state
is pure, presented by a (normalized) vector $e$. As we see from the formula
above (for updating the belief operator), the updated state is a probability
mixture (with weights $p_{i}$) of pure states corresponding to the
projections of $e$ on the subspaces $W_{i}$. That is a pure (coherent) state
disintegrates (decoheres) into a mixture of pure states.

\section{Quantum Cognition in Economics}

In this section we discuss the possible value of our results for behavioral
economics. We start with a few words about quantum cognition. We know that
in order to assess the world we build a representation of it, a "represented
world" which is a mental construct. In classical standard theory, the
represented world reflects our incomplete knowledge about the world
expressed in our beliefs and these beliefs (should) evolve according to
Bayes' rule in response to new information. Quantum cognition has been
developed under the last decades as alternative approach to incorporate two
observations: 1. People have difficulties to build a representations of a
complex object. What people do is to consider a complex object from
different perspectives - one at a time; 2. People may be unable to combine
perspectives i.e., to synthesize all relevant information into one stable
representation of the complex object.\ Quantum cognition models the
incapacity to combine some pieces of information by analogy with
incompatible properties (also called "Bohr complementary") in Quantum
Mechanics: different properties may be incompatible in the sense that they
cannot be given a determinate value simultaneously (cf speed and position)
but they complement each other in the description of the system. Similarly
different perspectives on an alternative may be incompatible in the mind in
the sense that the individual cannot have a clear stand with respect to them
simultaneously (i.e., combine them in a single coherent stable picture) but
the different perspectives contribute to characterizing the alternative.

In the present paper, we have extended decision theory to a non-classical
uncertainty environment. An interpretation of this move corresponds to
proposing that the "represented world" used to evaluate lotteries exhibits
quantum-like properties. In other words quantum indeterminacy of beliefs
captures the above mentioned cognitive limitations. Our results with respect
to dynamic consistency shed new light on observed behavioral anomalies in
the spirit\textbf{\ }of Shiller (2000)\ and Akerlof and Shiller (2015) who
write : "In our thoughts, as in our conversation, our minds may change. It
is not just that we acquire new "information"; we change our point of view
and we interpret information in a new way. Importantly these evolutions of
our thoughts mean that our opinions, and the decisions that are based on
them, may be quite inconsistent"(p.45). We next provide an example showing
how quantum indeterminacy of beliefs and in particular its dynamic
properties illustrated in Example 3 above delivers the kind of
manipulability at the core of Phishing for Phools.

Consider a seller who wants to sell a new smartphone at price 300\$ and a
customer considering buying one. Initially, the customer holds beliefs about
the quality of the smartphone that can be either Excellent (utility 600) or
Standard (utility 100). Her initial cognitive state\ assigns subjective
probabilities 0.25 and 0.75 respectively to the two possible events so the
expected utility is $EU(S)=.25\cdot 600+.75\cdot 100=225$. The alternative
is to keep the money which has utility 300. So initially the customer does
not want to buy the new smartphone since $225<300$. Now the seller engages
in a conversation about the use of the new smartphone among famous people.
In so doing the customer is moved from a private user perspective on the
smartphone to a "glamour" perspective (also with two outcomes: glamour, not
glamour). Assume that the two perspectives are incompatible in the mind. Now
what counts to our consumer is her idol Beyonc\'{e}, so she asks whether
Beyonc\'{e} uses that smartphone. The seller answers truthfully either yes
or no. After the conversation the consumer updates her beliefs, her
cognitive state is modified.\footnote{\textbf{One possible interpretation is
when in the "glamour" perspective, she is (possibly unconsciuosly) reminded
of the general disconnect between what is temporarily "fashionable" and
fundamental user values of commodities. }}\ Consider for simplicity the case
when private use and glamour perspectives are totally uncorrelated (the
corresponding bases are 45$%
{{}^\circ}%
$\ rotations of each other - see the figure below). As she turns back to the
question whether or not to buy the smartphone, the lottery with updated
beliefs yields $EU(S)=.5\cdot 600+.5\cdot 100=400\ >300$ whether she learned
that the smartphone is glamour or not (and her cognitive state is projected
onto the corresponding axis):"the phool has been phished" i.e., the seller
exploited the quantum indeterminacy of the consumer's "represented
smartphone" (that is her beliefs and her consistent updating rule when
evaluating the lottery) to change her preferences so she chooses to purchase
the smartphone. Quoting Akerlof and Shiller again " Just change people focus
and you can change the decisions they make" (p. 173).

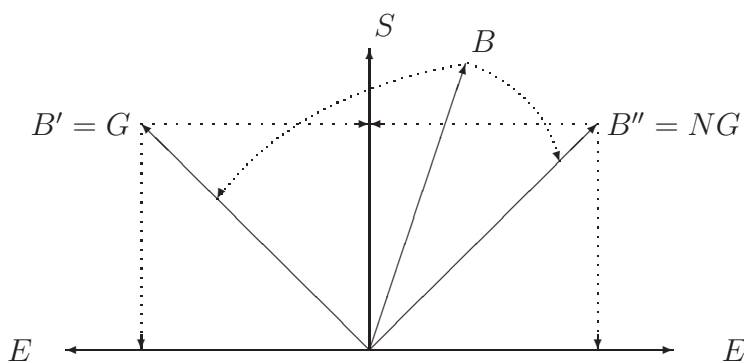
\unitlength=1mm \special{em:linewidth 0.4pt} \linethickness{0.4pt}
\begin{picture}(118.00,53.00)
\put(70.00,10.00){\vector(1,0){40.00}}
\put(70.00,10.00){\vector(-1,0){40.00}}
\put(70.00,10.00){\vector(0,1){40.00}}
\put(70.00,10.00){\vector(1,1){30.00}}
\put(70.00,10.00){\vector(-1,1){30.00}}
\put(70.00,10.00){\vector(1,3){12.67}}
\put(72.00,53.00){\makebox(0,0)[cc]{$S$}}
\put(85.00,51.00){\makebox(0,0)[cc]{$B$}}
\put(110.00,40.00){\makebox(0,0)[cc]{$B''=NG$}}
\put(118.00,10.00){\makebox(0,0)[cc]{$E$}}
\put(24.00,10.00){\makebox(0,0)[cc]{$E$}}
\put(32.00,40.00){\makebox(0,0)[cc]{$B'=G$}}
\bezier{24}(40.00,10.00)(40.00,24.00)(40.00,40.00)
\bezier{38}(40.00,40.00)(70.00,40.00)(100.00,40.00)
\bezier{24}(100.00,40.00)(100.00,25.00)(100.00,10.00)
\bezier{20}(83.00,48.00)(93.00,44.00)(95.00,35.00)
\bezier{40}(83.00,48.00)(60.00,44.00)(50.00,30.00)
\put(40.00,12.00){\vector(0,-1){2.00}}
\put(100.00,12.00){\vector(0,-1){2.00}}
\put(72.00,40.00){\vector(-1,0){2.00}}
\put(68.00,40.00){\vector(1,0){2.00}}
\put(94.00,38.00){\vector(1,-3){1.00}}
\put(51.00,32.00){\vector(-1,-2){1.00}}
\end{picture}

The idea of the Phishing equilibrium is that opportunities for manipulations
will be exploited whenever that is profitable. By recasting the standard
approach into non-classical uncertainty our theory can provide a rigorous
setting for investigating the properties of competitive markets with
manipulable consumers.

\section{Concluding remarks}

In this paper we provided a fully consistent choice theory integrating
cognitive limitations affecting our capacity to build representations of
choice alternatives. We found that the mathematical formalism of quantum
mechanics offers a suitable framework for modeling such cognitive
limitations. We fully characterize the rules for consistent choice behavior
in a non-classical uncertainty environment. A concept of quantum lottery is
introduced and sufficient and necessary conditions for choice behavior to be
representable by an expected utility function are formulated in a most
general setting. We also derive, from behavioral principles, an updating
rule that secures the dynamic consistency of the preference relation as the
decision-maker learns new information.

We found that most of the classical axioms of decision theory carry over to
the context of quantum lotteries. This is because all but one axiom can be
formulated in terms of a single orthogonal decomposition of the state space.
When considering a single orthogonal decomposition, quantum lotteries
operating in the Hilbert space are equivalent to roulette lotteries in a
classical state space. An additional axiom is required to secure that the
probability for any specific event does not depend on the particular lottery
that it belongs to. The necessity to impose that axiom stems from the fact
that while it is trivially true in the classical world, it is not necessary
so in our general setting.

A most interesting result is that the von Neumann-L\"{u}ders postulate which
is central to Quantum Mechanics and informs about the impact of a
measurement on the state of a system can be derived from a consistency
requirement on choice behavior. When the belief-state (cognitive state) is
updated according to the postulate, the agent conditional preferences
reflect a single preference order. Of particular interest for behavioral
economics is that in contrast with classical subjective expected utility
theory, dynamic consistency of preferences does not entail the so-called
recursive dynamic consistency. This is an expression of the fundamental
distinction between the two settings namely that the resolution of
uncertainty depends on the operation(s) performed to resolve it. We suggest
that this very feature makes our theory an attractive candidate to develop
"The Economics of Manipulation and Deception" called for by Akerlof and
Shiller.

\section*{Appendix 1. Elementary facts about Hilbert spaces}

\emph{Hilbert space}

Let $\mathbb{R}$ and ${\mathbb{C}}$ denote the fields of real and of complex
numbers. For a complex number $z$, $\bar{z}$ denotes the complex conjugate
number.\medskip

\textbf{Definition 1.} Let $H$ be a vector space over the field $\mathbb{C}$%
. An \emph{Hermitian form} on $H$ is a mapping $(.,.):H\times H\rightarrow
\mathbb{C}$ such that: a) it is linear in the first argument; b) $(v,w)=%
\overline{(w,v)}$ for any $v,w\in H\ $(in particular, $(v,v)$ is a real
number); c) $(v,v)\geq 0$ for any $v\in H$, and $(v,v)=0$ only for $v=0$%
.\medskip

Vectors $v$ and $w$ are called \emph{orthogonal} if $(v,w)=0$; in this case $%
(w,v)=0$ as well.

A \emph{Hilbert space} is a vector space $H$ endowed with a Hermitian form,
which is complete relatively to the norm $|v|=\sqrt{(v,v)}$. In order to
avoid unnecessary difficulties and subtleties we assume further that $H$ has
finite dimension; then $H$ automatically is complete.\medskip

When discussing lotteries and measurements we shall not be dealing so much
with vectors in $H$ as with special operators (linear mappings from $H $ to $%
H$) called Hermitian operators. 
\medskip

\textit{Hermitian operators}

\textbf{Definition 2. } A linear operator $A:H\rightarrow H$ is called \emph{%
Hermitian}, if $(Av,w)=(v,Aw)$ for any $v,w\in H$.\medskip

Clearly $\left( Av,v\right) $ is a real number for any $v\in H$. Hermitian
operator $A$ is called \emph{nonnegative} if $\left( Av,v\right) \geq 0$ for
any $v$. For Hermitian operators $A$ and $B$ we write $A\geq B$ if $A-B$ is
nonnegative. The identity operator $E$ ($Ev=v$ for every $v\in H$) is
Hermitian.

A most important, for the purpose of this work, class of Hermitian operator
consists of projectors. A projector is an idempotent Hermitian operator,
that is $PP=P$. Since $(Pv,v)=(PPv,v)=(Pv,Pv)\geq 0$, any projector is
nonnegative. Each projector $P$ define a vector subspace $V=\text{Im}%
P\subset H$, consisting of vectors $v$ such that with $Pv=v$. The kernel of
the projector consists of vectors orthogonal to $V$, $\text{Ker}P=V^{\perp }$%
. The set of projectors can be identify with the set of (closed) subspaces
of $H$.

Any linear combination of Hermitian operators with real coefficients is an
Hermitian operator. In other words, the set $\mathbf{Herm}(H)$ of Hermitian
operators is a real vector space. The crucial importance of projectors is
underlined by the following important theorem.\medskip

\textbf{Spectral theorem}. \emph{Let $A$ be a Hermitian operator. Then there
exists a family of projectors $P_{i}$ and real numbers $a_{i}$ such that: 1)
$P_{i}P_{j}=0$ for $i\neq j$, 2) $\sum _i P_i=E$, and 3) $A=\sum_i a_i P_{i}$%
.} \medskip

In other words, in some orthogonal basis the operator $A$ can be represented
by a diagonal matrix (with real coefficients). The coefficients $a_i$ are
eigenvalues of the operator $A$. The set of numbers $a_i$ is called the
\emph{spectrum} of the operator $A$. Clearly $A$ is nonnegative if and only
if all coefficients $a_{i}$ are nonnegative. An operator $A$ is a projector
if and only if its spectrum $Spec A$ consists of 0 or 1. \medskip

Each nonnegative operator $A$ has a (unique) nonnegative square root $\sqrt{A%
}$ (or $A^{1/2}$, such that $(\sqrt{A})^2=A$). If $A=\sum_i a_i P_{i}$ is a
spectral representation of $A$ then $\sqrt{A}=\sum_i \sqrt{a_i} P_{i}$%
.\medskip

\emph{Trace of operators}

For arbitrary (not necessarily Hermitian) linear operator $A:H\rightarrow H$
it is possible to talk about its trace $\mathbf{Tr}\left( A\right) $. More
precisely, for any quadratic matrix $A=(a_{ij})$, the trace $\mathbf{Tr}(A)$
is defined as $\sum_i a_{ii}$, the sum of its diagonal elements. A
remarkable property of the trace is its `commutativity': $\mathbf{Tr}(AB)=%
\mathbf{Tr}(BA)$ for any quadratic matrix $A$ and $B$. This in particular
implies that the trace of an operator is independent of the choice of basis,
thereby allowing for an unambiguous definition of the trace of a linear
operator.

For example, $\mathbf{Tr}(E)=\dim H$. More general, if $P$ is an
(orthogonal) projector (on subspace $V=ImP$) then $\mathbf{Tr}(P)=\dim V$.
Due to the spectral theorem, we obtain that the trace of Hermitian operator $%
A=\sum a_{i}P_{i}$ is equal to $\sum a_{i}\text{rk}(P_{i})$ and, in
particular, it is a real number. The trace of nonnegative operator $A$ is
nonnegative and is strictly positive if $A\neq 0$.

For two Hermitian operators $A$ and $B$ define $(A,B)_{Herm}=\mathbf{Tr}(AB)$%
. We assert that this `scalar product' gives a structure of Euclidean space
on the real vector space $\mathbf{Herm}(H)$. This follows from Lemma
below.\medskip

\textbf{Lemma.} a) $(A,B)_{Herm}$ \emph{is a real number};

b) $(A,B)_{Herm}=(B,A)_{Herm}$;

c) $(A,A)_{Herm}\ge 0$ \emph{and is equal to $0$ only if} $A=0$. \medskip

\emph{Proof.} Due to the `commutativity' of the trace, $2\mathbf{Tr}(AB)=%
\mathbf{Tr}(AB)+\mathbf{Tr}(BA)=\mathbf{Tr}(AB+BA)$. It is easy to
understand that the operator $AB+BA$ is Hermitian, hence its trace is real.
This proves a).

b) follows from the `commutativity' of the trace.

c) follows from the nonnegativity of the operator $A^{2}$. $\Box $

\section*{Appendix 2. Proof of Theorem 1}

Let $\preceq $ be a nice preference relation on the set $\mathbf{QL}(H)$. We
shall be working with roulette-valued Q-lotteries, that is with expressions
of the form $\sum_{i}l_{i}\otimes P_{i}$, {where $(P_{i},\ i\in I)$ is a
PDU, and $(l_{i},\ i\in I)$ is a collection of roulette lotteries on $X$}. {%
Let $\mathbf{QL}_\mathcal{P}(H,\Delta(X))$ denote the set of all roulette
valued Q-lotteries with base $\mathcal{P}$}. We first provide a result that
shows that mixtures of {canonical Q-lotteries in $\mathbf{QL}(H)$ and
mixtures of Q-lotteries in $\mathbf{QL}_\mathcal{P}(H,\Delta(X))$ are
\textbf{compatible}.}

Consider a lottery in the canonnical form $\sigma=\sum_{x}x\otimes P_{x},$
where $x\in X,P_{x}$ are nonnegative Hermitian operators which add up to $E,$
$\sum_{x}P_{x}=E.$\ The mixture $\alpha \sigma +\left( 1-\alpha \right) \tau
, $ where $\tau =\sum_{x}x\otimes Q_{x}$ {and $\alpha \in [0,1]$}, is given
as $\sum_{x}x\otimes \left( \alpha P_{x}+\left( 1-\alpha \right)
Q_{x}\right) .$\ On the other hand a roulette valued Q-lottery writes $%
\sigma=\sum_{i}l_{i}\otimes P_{i}$, {where $\mathcal{P}=(P_{i},i\in I)$} is
a measurement device (the base of the lottery) and $l_{i}\in \Delta \left(
X\right) .$ A mixture of such lotteries is defined by the following formula:
$\alpha \sigma+(1-\alpha)\tau=\sum_{i}\left( \alpha l_{i}+\left( 1-\alpha
\right) r_{i}\right) P_{i}$, {where $\tau=\sum_{i}r_{i}\otimes P_{i}$ is
another Q-lottery in $\mathbf{QL}_\mathcal{P}(H,\Delta(X))$ and $\alpha \in
[0,1]$}.

We next define the canonisation mapping $can:\mathbf{QL}_\mathcal{P}
\rightarrow \mathbf{QL}_c,$ which maps lottery $\sum_{i}l_{i}\otimes P_{i}$
(where the roulette lottery $l_{i}$ has the form $\sum_{x}x\otimes
l_{i}\left( x\right) $ i.e., $l_{i}$ gives value $x$ with probability $%
l_{i}\left( x\right) )$ into the canonical lottery $\sum_{x}x\otimes \left(
\sum_{i}l_{i}\left( x\right) P_{i}\right) .\medskip $

\textbf{Lemma 2. } {\emph{The mapping} can \emph{preserves the operation of
mixture; that is, for any $\sigma,\tau \in \mathbf{QL}_\mathcal{P}%
(H,\Delta(X)) $ and $\alpha \in [0,1]$, $can(\alpha \sigma +\left( 1-\alpha
\right) \tau) =\alpha can\left( \sigma \right) +\left( 1-\alpha \right)
can\left( \tau \right)$}.}

\textit{Proof}. Assume we have two $\mathcal{P-}$based lotteries $\sigma
=\sum_{i}l_{i}\otimes P_{i}$ and $\tau =\sum_{i}r_{i}\otimes P_{i}$ and some
$\alpha \in \left[ 0,1\right].$ We want to show that {$can(\alpha \sigma
+\left( 1-\alpha \right) \tau) =\alpha can\left( \sigma \right) +\left(
1-\alpha \right) can\left( \tau \right). $}

The left hand side is equal to $can\left( \sum_{i}\left( \alpha l_{i}+\left(
1-\alpha \right) r_{i}\right) \otimes P_{i}\right) =$

$\sum_{x}x\otimes \sum_{i}\left( \alpha l_{i}+\left( 1-\alpha \right)
r_{i}\right) \left( x\right) P_{i}=\sum_{x}x\otimes \sum_{i}\left( \alpha
l_{i}\left( x\right) +\left( 1-\alpha \right) r_{i}\left( x\right) \right)
P_{i}$ $=$

$=\sum_{x}x\otimes \sum_{i}\left( \alpha \sum_{i}l_{i}\left( x\right)
+\left( 1-\alpha \right) \sum_{i}r_{i}\left( x\right) \right) P_{i}.$

{The right hand side is $\alpha \sum_{x}x\otimes \left( \sum_{i}l_{i}\left(
x\right) P_{i}\right) +\left( 1-\alpha \right) \left( \sum_{x}x\otimes
\sum_{i}r_{i}\left( x\right) P_{i}\right) $}

{$=\alpha can\left( \sigma \right)+\left( 1-\alpha \right) can\left( \tau
\right)$. $\Box $} \medskip \medskip

Returning to the proof of Theorem 1,\ let $\preceq _{\Delta }$ denote the
derived preference relation on the set $\Delta (X)$ of roulette lotteries.
The assertion of the theorem is true if the preference $\preceq $ is
trivial. Indeed, we can take $u$ to be a constant and take an arbitrary
functional $\beta $. So, from now on, we assume that the preference $\preceq
$ is nontrivial. That is $\tau \prec \sigma $ for some Q-lotteries $\sigma
=\sum_{i}l_{i}\otimes P_{i}$ and $\tau =\sum_{j}m_{j}\otimes Q_{j}$.\medskip

\textbf{Claim 0.} \emph{Let $l^*$ be the best lottery among $(l_i)$ with
respect to the derived weak order $\preceq _\Delta $. Then $\sigma \preceq
l^*\otimes E$. }\medskip

Proof. Consider the `constant' lottery $\sigma ^*=\sum _i l^*\otimes P_i$.
Due to A4, $\sigma \preceq \sigma ^*$. Due to A0, $\sigma ^*\approx
l^*\otimes E$. Due to transitivity of $\preceq $ (see A1) we conclude that $%
\sigma \preceq l^*\otimes E$. $\Box$\medskip

\textbf{Corollary 1.} \emph{The derived preference $\preceq _{\Delta }$ is
non-trivial, that is $l_{\ast }\prec _{\Delta }l^{\ast }$ for some ordinary
lotteries $l_{\ast }$ and $l^{\ast }$.}\medskip

{Proof.} Indeed, if $m_*$ ia the worst lottery among $(m_j)$, then we have
\begin{equation*}
m_*\otimes E\preceq \tau \prec \sigma \preceq l^*\otimes E,
\end{equation*}
{whence} $m_*\prec _\Delta l^*$. $\Box$\medskip

We fix such lotteries $l_*\prec _\Delta l^*$; a function $u:\Delta (X)\to
\mathbb{R }$ is said to be \emph{normalized} if $u(l_*)=0$ and $u(l^*)=1$.

Fix now some measurement device $\mathcal{P }=(P_i, \ i\in I)$, and {let $%
\mathbf{QL}_\mathcal{P}(H,\Delta(X))$ denote the set of all roulette valued
Q-lotteries with base $\mathcal{P}$. We first extend $\preceq$ into a
preference relation $\preceq_\mathcal{P}$ defined on $\mathbf{QL}_\mathcal{P}%
(H,\Delta(X))$ by setting $\sigma \preceq \tau$ if and only if $%
can(\sigma)\preceq can(\tau)$ for any $\sigma,\tau \in \mathbf{QL}_\mathcal{P%
}(H,\Delta(X))$}. \medskip

\textbf{Claim 1.} \emph{There exists a normalized affine function $u_{%
\mathcal{P}}$ on $\Delta (X)$ and a function $\beta _{\mathcal{P}}$ on the
set of outcomes $I$ ($\beta _{\mathcal{P}}(i)\geq 0$ and $\sum_{i}\beta _{%
\mathcal{P}}(i)=1$) such that the preference $\preceq _{\mathcal{P}}$ is
represented by the function $U_{\mathcal{P}}$, $U_{\mathcal{P}%
}(\sum_{i}l_{i}\otimes P_{i})=\sum_{i}u_{\mathcal{P}}(l_{i})\beta _{\mathcal{%
P}}(i)$. Moreover, both $u_{\mathcal{P}}$ and $\beta _{\mathcal{P}}$ are
unique. }\medskip

Proof. Each Q-lottery $\sigma =\sum_{i}l_{i}\otimes P_{i}$ can be considered
as a `horse' lottery $f:I\rightarrow \Delta (X)$, where $f(i)=l_{i}$.
Moreover, due to axioms A1-A4 and Lemma 2, the relation $\preceq _{\mathcal{P%
}}$ satisfies all the Anscombe-Aumann axioms. Therefore, by theorem 13.2 in
Fishburn (1970), we obtain an affine utility function $u_{\mathcal{P}}$ on $%
\Delta (X)$ and a probability measure $\beta _{\mathcal{P}}\in \Delta (I)$
that achive the representation stated in Claim 1. The uniqueness of $\beta _{%
\mathcal{P}}$ is also given by this theorem. The uniqueness of $u_{\mathcal{P%
}}$ follows from normalization of $u_{\mathcal{P}}$. $\Box $\medskip

\textbf{Claim 2.} \emph{The {functions $u_\mathcal{P }$ are} independent of $%
\mathcal{P }$ (and we denote them as $u$). }\medskip

Proof. Due to A0, each of the functions $u_{\mathcal{P}}$ represents the
derived preference $\preceq _{\Delta }$ on $\Delta (X)$. Therefore they are
positive affine transformations of each other. Normalization gives that they
are in fact equal to each other. $\Box $\medskip

\textbf{Claim 3.} \emph{For any Q-lottery $\sigma $, there exists an
ordinary lottery $l\in \Delta (X)$ {such that $\sigma $ is equivalent to $l$%
, that is} $\sigma \approx l\otimes E$.}\medskip

Proof. Let $\sigma =\sum _i l_i\otimes P_i$, and $l^b$ (correspondingly, $%
l^w $) is a best (corresp., a worst) lotteries among $(l_i, i\in I)$. Due to
A4, we have $\sum _i l^w\otimes P_i \preceq \sigma \preceq \sum _i
l^b\otimes P_i $, and all these Q-lotteries have the same base $\mathcal{P }%
=(P_i, i\in I)$. Therefore we can apply Claim 1, which gives inequalities
\begin{equation*}
u(l^w)\le \sum _i u(l_i) \beta _\mathcal{P }(i) \le u(l^b).
\end{equation*}
Hence $\sum _i u(l_i)\beta _\mathcal{P }(i)=\alpha u (l^w) +(1-\alpha )
u(l^b)=u(l)$ for some $\alpha \in [0,1]$, where $l=\alpha l^w+(1-\alpha )l^b$%
. By Claim 1, we have $\sigma \approx \sum _i l\otimes P_i \approx l\otimes
E $. $\Box$\medskip

Due to Claim 1, the function $U_\mathcal{P }$ allows to compare Q-lotteries
with base $\mathcal{P }$. But we assert that it allows to compare {%
Q-lotteries} with different {bases} as well.\medskip

\textbf{Claim 4.} \emph{Let $\sigma =\sum _i l_i\otimes P_i$ be a Q-lottery
with a base $\mathcal{P }=(P_i, i\in I)$, and let $\tau =\sum _j m_j\otimes
Q_j$ be a Q-lottery with a base $\mathcal{Q }=(Q_j, j\in J)$. Then $\sigma
\preceq \tau $ if and only if $U_\mathcal{P }(\sigma )\le U_\mathcal{Q }%
(\tau )$. }\medskip

Proof. Due to Claim 3, the lottery $\sigma $ is equivalent to some lottery $%
l\otimes E$, or to the lottery $\sum _i l\otimes P_i$. Therefore, $U_%
\mathcal{P }(\sigma )=U_\mathcal{P }(\sum _i l\otimes P_i)=u(l)$. Similarly,
$\tau $ is equivalent to $m\otimes E$, and $U_\mathcal{Q }(\tau )=u(m)$. Now
\begin{equation*}
\sigma \preceq \tau \Leftrightarrow l\otimes E \preceq m\otimes E
\Leftrightarrow U_\mathcal{P }(\sigma )=u(l)\le u(m)=U_\mathcal{Q }(\tau ).
\  \  \  \Box
\end{equation*}

Let us return now to the functions $\beta _\mathcal{P }$. We assert that $%
\beta _\mathcal{P }(i)$ depends only on the operator $P_i$, not of $\mathcal{%
P } $ and $i$.\medskip

\textbf{Claim 5.} \emph{Let $\mathcal{P }=(P_1,...,P_n)$ and $Q=(Q_1, ...
Q_k)$ be two measurement devices (bases), and $P_1=Q_1=R$. Then $\beta _%
\mathcal{P } (1)=\beta _\mathcal{Q }(1)$.}\medskip

Proof. Consider Q-lottery $\sigma =l^*\otimes P_1+\sum _{i=2}^n l_*\otimes
P_i$ with base $\mathcal{P }$. Its $\mathcal{P }$-utility $U_\mathcal{P }%
(\sigma )$ is equal to $\beta _\mathcal{P }(1)$. Now let us form the
auxiliary base $\mathcal{R }=(R, E-R)$ and the following Q-lottery $\rho
=l^*\otimes R+l_*(E-R)$. Since $E-R=P_2+...+P_n$, the lottery $\rho $ is
equivalent to $\sigma $ (see Axiom A0). Therefore $\mathcal{R }$-utility $U_%
\mathcal{R }(\rho )$ (which equals $\beta _\mathcal{R }(1)$) is, by Claim 4,
equal to $\beta _\mathcal{P }(1)$. The same applies to $\mathcal{Q }$ and
gives the equality $\beta _\mathcal{R }(1)=\beta _\mathcal{Q }(1)$. Together
with the equality $\beta _\mathcal{R }(1)=\beta _\mathcal{P }(1)$ we obtain
the equality $\beta _\mathcal{P }(1)=\beta _\mathcal{Q }(1)$. $\Box$\medskip

As a consequence, we can speak about the number $\beta (P)$ for any `event' $%
P$, {that is} for any hermitian operator $P$, $0\le P\le E$. $\beta (P)$ is $%
\mathcal{R }$-utility $U_\mathcal{R }$ of the following Q-lottery $%
l^*\otimes P+l_*\otimes (E-P)$. Correspondingly, the utility of an arbitrary
Q-lottery $\sigma =\sum _i l_i\otimes P_i$ can be rewritten as
\begin{equation*}
U(\sigma )=\sum _i u(l_i)\beta (P_i).
\end{equation*}

Obviously, $\beta (0)=0$ and $\beta (E)=1$. Moreover, $\beta (P)\ge 0$ for
any `event' $P$, and $\sum _i \beta (P_i)=1$ provided $\sum _i P_i=E$%
.\medskip

\textbf{Claim 6.} \emph{If $0\le P$, $0\le Q$, and $P+Q\le E$, then $\beta
(P+Q)=\beta (P)+\beta (Q)$.}\medskip

{Proof.} Indeed, consider the Q-lottery {$l^*\otimes P+l^*\otimes
Q+l_*\otimes (E-P-Q)$}. Its utility is $\beta (P)+\beta (Q)$. On the other
hand, due to A0, this lottery is equivalent to the lottery $l^*\otimes
(P+Q)+l_*\otimes (E-P-Q)$, whose utility is $\beta (P+Q)$. $\Box$\medskip

{Claim 6} implies that $\beta $ can be extended to a (unique) linear
functional $\beta $ on the vector space $\mathtt{Herm}(H)$. Obviously, $%
\beta (A)\geq 0$ for $A\geq 0$, and $\beta (E)=1$. That is $\beta $ is a
belief functional. This completes the proof of Theorem 1.

\section*{Appendix 3. Proof of Theorem 2}

Proof of Assertion a). Here we can work with lotteries in the form of
Hermitian operators. Utility $U(A)$ of such an operator $A$ is equal to $%
\mathbf{Tr}(AB)$ and utility under the condition $P$ is equal to $U(A|P)=%
\mathbf{Tr}(APBP)$ (up to the factor $\mathbf{Tr}(PBP)$).

Proving property \textbf{A5}. We assume that $A|W=0$ and have to show that $%
U(A|P)=0$. Note that $A|P=0$ is equivalent to $PAP=0$. Now $U(A|P)=\mathbf{Tr%
}(APBP)=\mathbf{Tr}(PAPB)=\mathbf{Tr}(0B)=0$.

Proving property \textbf{A6}. Here we assume that $A$ commute with $P$ and
that $(E-P)A=0$ (that is $A=PA=AP$). We have to show that $U(A|P)$ is equal
to $U(A)$. But $U(A|P)=\mathbf{Tr}(APBP)=\mathbf{Tr}(PAPB)=\mathbf{Tr}%
(AB)=U(A)$, because $PAP=A$.\medskip

Proof of Assertion b). To proof it, we define (for an arbitrary Q-lottery $%
\sigma =\sum_{i}x_{i}\otimes P_{i}$ and an event $W$ given by a projector $P$%
) some special (\emph{adapted to} $P$) Q-lottery that we denote $%
ad_{W}(\sigma )$. Roughly speaking, $ad_{W}(\sigma )=\sum_{i}x_{i}\otimes
PP_{i}P$. However, the sum $\sum_{i}PP_{i}P$ is equal to $P$, not to $E$.
Therefore we add to this sum a tail-end $x_{\ast }\otimes (E-P)$. Here $%
x_{\ast }$ is a prize with zero utility. The final formula yields
\begin{equation*}
ad_{W}(\sigma )=\sum_{i}x_{i}\otimes PP_{i}P+x_{\ast }\otimes (E-P).
\end{equation*}

{\textbf{Lemma 3. }} a) $\sigma |W=ad_{W}(\sigma )|W$;

b) \emph{the adapted lottery $ad_W(\sigma)$ is compatible with $P$; }

c) \emph{for any Q-lottery} $\sigma $ \emph{we have} $ad_{W}(\sigma
)|W^{\perp }=x_{\ast }\otimes (E-P)$.\medskip

{Proof.} a) It is obvious that $P_{i}|W=PP_{i}P|W$. Moreover, $%
(E-P)|W=P(E-P)=0$.

b) For any $i$ we have $PPP_iP=PP_iPP$, since $PP=P$. Moreover, $%
P(E-P)=(E-P)P=0$.

c) It is clear that $(E-P)PP_iP=0$ for any $i$. $\Box$\medskip

\textbf{Proposition 4.} \emph{Suppose that a preference relation $\preceq
_{W}$ on the set $\mathbf{QL}(H)$ is a weak order and satisfies the axioms
A5 and A6. Then it is given by the following explicit formula (where $\sigma
$ and $\tau $ are Q-lotteries on $H$):}
\begin{equation}
\sigma \preceq _{W}\tau \text{ if and only if }ad_{W}(\sigma )\preceq
ad_{W}(\tau ).  \label{up}
\end{equation}

Indeed, due to the axiom A5 and {Lemma 3}, we have $\sigma \approx
_{W}ad_{W}(\sigma )$ and $\tau \approx _{W}ad_{W}(\tau )$. Applying axiom A6
to the lotteries $ad_{W}(\sigma )$ and $ad_{W}(\tau )$ (which is possible
due to points 2) and 3) of {Lemma 3}), we obtain that $ad_{W}(\sigma
)\preceq ad_{W}(\tau )$ if and only if $ad_{W}(\sigma )\preceq
_{W}ad_{W}(\tau )$. The rest follows from the transitivity of $\preceq _{W}$%
. $\Box $\medskip

{There remains} to recall that (unconditional) utility of lottery $%
ad_W(\sigma)$ is equal to {$\mathbf{Tr}(Sh(ad_W(\sigma))B)$}. If $%
A=Sh(\sigma)$ then {$Sh(ad_W(\sigma))=PAP+0(E-P)=PAP$}. Therefore {$%
U(ad_W(\sigma))=\mathbf{Tr}(PAPB)=\mathbf{Tr}(APBP)=\mathbf{Tr}(PBP)\mathbf{%
Tr}(AB^{up})$}, what is (up to the factor $\mathbf{Tr}(PBP)$) the utility of
$\sigma$ with respect to `updated' belief operator $B^{up}=PBP/\mathbf{Tr}%
(PBP)$. {This completes the proof of Theorem 2.}

\section*{Appendix 4. Proof of Proposition 1.}

We here prove an assertion that generalizes Propostion 1 as we substitute
projector $P$ with an arbitrary `fuzzy-projector' $F$, that is an operator $%
0\leq F\leq E$ (see Remark 4 of Section 4). The posterior $B^{up}$ is given
as $\sqrt{F}B\sqrt{F}/\mathbf{Tr}(\sqrt{F}B\sqrt{F})=\sqrt{F}B\sqrt{F}/%
\mathbf{Tr}(FB)$. We shall denote it as $B|F$.\medskip

\textbf{Proposition $1^{\prime }$.} \emph{The distance between a prior $B$
and the posterior $B|F$ is} $O((1-\mathbf{Tr}(FB))^{1/4})$.\medskip

In other words, if $\varepsilon =1-\mathbf{Tr}(FB)$ then the distance
between $B$ and $B|F$ is $O(\varepsilon ^{1/4})$.\medskip

{Proof.} Choose an orthonormal basis of $H$ in which the operator $F$ (as
well as $\sqrt{F}$) has diagonal form
\begin{equation*}
F=\mathtt{diag}(f_{1},...,f_{n}).
\end{equation*}%
In this basis operator $B$ is represented by Hermitian matrix $(b_{ij})$,
where $i$ and $j$ run over 1 to $n=\dim H$. The matrix $\sqrt{F}B\sqrt{F}$
has coefficients $\sqrt{f_{i}}b_{ij}\sqrt{f_{j}}$. The matrix $\sqrt{F}B%
\sqrt{F}$ differs of $B|F=\sqrt{F}B\sqrt{F}/Tr(FB)$ by less than $%
O(\varepsilon )$. Thus we need to compare $B$ and $\sqrt{F}B\sqrt{F}$ and to
show that the distance between them is $O(\varepsilon ^{1/4})$. Or
equivalenetly we need to show that (for any $i$ and $j$) distance between $%
b_{ij}$ and $\sqrt{f_{i}}b_{ij}\sqrt{f_{j}}$ is $O(\varepsilon ^{1/4})$ .

Let us divide the set of indices $i$ from $\{1,...,n\}$ into two groups. Say
that an index $i$ is \emph{non-essential}, if $b_{ii}\leq \sqrt{\varepsilon }
$, and is \emph{essential} in the opposite case.\medskip

{\textbf{Lemma 4. }} \emph{If $i$ is non-essential then $|b_{ij}|\leq
\varepsilon ^{1/4}$.}\medskip

{Proof.} A sub-matrix of the matrix $B$, formed by the rows and columns from
the set $\{i,j\}$, is Hermitian and therefore has non-negative determinant $%
b_{ii}b_{jj}-|b_{ij}|^2$. That is $\sqrt{\varepsilon }\ge b_{ii}b_{ij}\ge
|b_{ij}|^2$. $\Box$\medskip

Due to Lemma A, if $i$ or $j$ is non-essential then the coefficient $b_{ij}$
of the matrix $B$ as well as the corresponding coefficient $b_{ij}\sqrt{f_i}%
\sqrt{f_j}$ of the matrix $\sqrt{F}B\sqrt{F}$ is $\le \varepsilon ^{1/4}$.
Therefore the distance between them is $\le 2\varepsilon ^{1/4}$.

Hence we can suppose that the both indexes $i$ and $j$ are essential.\medskip

{\textbf{Lemma 5.}} \emph{If an index $i$ is essential then $%
f_i=1+O(\varepsilon ^{1/2})$.}\medskip

{Proof.} We have $Tr(FB)=1-\varepsilon $, that is $\sum _i
f_ib_{ii}=1-\varepsilon $. Moreover the trace of $B$ is equal to 1, that is $%
\sum _i b_{ii}=1$. Subtracting we obtain that $\sum _i
b_{ii}(1-f_i)=\varepsilon $, therefore each term of this sum is less or
equal to $\varepsilon$, $b_{ii}(1-f_i)\le \varepsilon $. . In particular, if
$i$ is essential then $b_{ii}>\varepsilon ^{1/2}$ and $1-f_i<\varepsilon
^{1/2}$, that is $f_i=1+O(\varepsilon ^{1/2})$. $\Box$\medskip

As a corollary we obtain (for an essential $i$) that $\sqrt{f_i}%
=1+O(\varepsilon^{1/2})$.

Let us return to evaluation of the distance between $b_{ij}$ and $b_{ij}%
\sqrt{f_{i}}\sqrt{f_{j}}$ in the case when $i$ and $j$ are essential
indexes. It is clear that $b_{ij}-b_{ij}\sqrt{f_{i}}\sqrt{f_{j}}%
=b_{ij}(1-(1+O(\varepsilon ^{1/2}))(1+O(\varepsilon ^{1/2})))$. The module
of this number is no more than $|b_{ij}|$ (which is $\leq 1$) multiplied by $%
Q(\varepsilon ^{1/2})$. Therefore the distance is no more than $\leq
O(\varepsilon ^{1/2})\leq O(\varepsilon ^{1/4})$ which proves Proposition $%
1` $. {$\Box$}\medskip

\end{document}